\documentclass[aps,prb,reprint,superscriptaddress,floatfix]{revtex4-2}
\usepackage{graphicx} 
\usepackage{xcolor}
\usepackage{amsmath} 
\usepackage{amssymb}
\usepackage{hyperref} 
\usepackage{caption}
\usepackage{subcaption}

\usepackage{amsmath,amsfonts,amssymb}
\DeclareMathAlphabet{\mathbbold}{U}{bbold}{m}{n}

\date{\today}

\begin{document}

\newcommand{\karim}[1]{\color{blue} #1 \color{black}}

\title{Mixture of Experts Framework in Machine Learning Interatomic Potentials for Atomistic Simulations}

\author{Gabriel de Miranda Nascimento}
\affiliation{Department of Materials Science and Engineering, Massachusetts Institute of Technology, Cambridge, MA, USA}
\affiliation{John A. Paulson School of Engineering and Applied Sciences, Harvard University, Cambridge, MA, USA}

\author{Marc L. Descoteaux}
\author{Laura Zichi}
\author{Chuin Wei Tan}
\author{William C. Witt}
\affiliation{John A. Paulson School of Engineering and Applied Sciences, Harvard University, Cambridge, MA, USA}

\author{Nicola Molinari}
\author{Sriteja Mantha}
\author{Daniil Kitchaev}
\author{Mordechai Kornbluth}
\author{Karim Gadelrab}
\author{Charles Tuffile}
\affiliation{Robert Bosch LLC Research and Technology Center, Watertown, MA, USA}

\author{Boris Kozinsky}
\affiliation{John A. Paulson School of Engineering and Applied Sciences, Harvard University, Cambridge, MA, USA}
\affiliation{Robert Bosch LLC Research and Technology Center, Watertown, MA, USA}

\begin{abstract}
    First-principles atomistic simulations are essential for understanding complex material phenomena but are fundamentally limited by their computational cost. While Machine Learning Interatomic Potentials (MLIPs) have drastically improved cost for a given accuracy, their inference cost remains a bottleneck for massive systems or long timescales. To address this, we introduce a multifidelity ``Mixture-of-Experts'' framework based on the E(3)-equivariant \textit{Allegro} architecture. Our method spatially partitions the simulation domain into a chemically complex region (e.g., reactive interfaces) and a simple region (e.g., bulk lattice), assigning models of varying capacity to each. Among the challenges in such static domain decomposition, the mechanical mismatch between models at the interface is particularly critical, as it can generate artificial stress fields and instability. We address this challenge with a co-training strategy in which the loss function includes agreement constraints---penalties on per-atom energy and force discrepancies between models evaluated on shared bulk environments---forcing the independent models to learn a consistent physical description of the bulk material. We validate this approach on a realistic Pt+CO catalytic system, demonstrating that the co-trained models maintain exact energy conservation, align their bulk mechanical response (e.g., equation of state and bulk modulus), and achieve predictive accuracy comparable to a full high-fidelity simulation at more than twice the computational speed.

\end{abstract}

\maketitle

\section{Introduction}

The central challenge in computational materials modeling remains the steep trade-off between predictive accuracy and computational scalability. While first-principles methods like Density Functional Theory (DFT) offer quantum-mechanical rigor, their computational cost typically limits them to small systems. Machine Learning Interatomic Potentials (MLIPs), particularly those based on E(3)-equivariant architectures~\cite{BatatiaEtAlDesignSpaceE3Equivariant2022, batzner20223, NEURIPS2022_4a36c3c5, bochkarev2024graph}, have emerged as a transformative solution, achieving near-DFT accuracy at a fraction of the cost. However, the inference cost of deep equivariant models remains orders of magnitude higher than classical force fields, rendering the simulation of massive systems or long-timescale phenomena prohibitively expensive for a single high-fidelity model. This creates a critical need for strategies that can intelligently allocate computational resources where they are chemically most needed.

To address this, we propose a multifidelity framework inspired by classical QM/MM methods~\cite{QMMM, bernstein2009hybrid} and "Mixture-of-Experts" (MoE) architectures in deep learning~\cite{sun2024mixture, wang2025hmoe}. Just as heterogeneous MoE models route information-dense tokens to high-capacity experts, our approach spatially partitions the simulation domain based on chemical complexity. We assign a ``high-fidelity'' model to chemically complex regions (e.g., reactive catalytic surfaces) and a computationally cheaper ``low-fidelity'' model to simpler environments (e.g., bulk lattice) that, while less chemically rich, are nonetheless necessary boundary conditions for a physically meaningful simulation. By treating the local atomic environment as the routing context, the high-fidelity computational cost scales with the size of the chemically complex region rather than the total number of atoms, optimizing the cost-accuracy Pareto frontier.

Recent efforts have sought to operationalize this multifidelity concept through two primary strategies: force-mixing and energy-mixing. The force-mixing approach, exemplified by the ML-MIX package (Birks et al.~\cite{birks2025efficient}), blends atomic forces from different potentials over a spatial buffer region. While this method allows for flexible coupling of heterogeneous models (e.g., ACE~\cite{LysogorskiyEtAlPerformantImplementationAtomic2021} and UF3~\cite{XieEtAlUltrafastInterpretableMachinelearning2023}) via constrained linear fitting of elastic constants, the blended forces are not derivable from a single underlying potential and the spatially varying mixing weights break global momentum conservation. The resulting non-conservative dynamics may exhibit energy drift, necessitating the use of thermostats to stabilize the simulation, which precludes true Hamiltonian dynamics in the microcanonical ($NVE$) ensemble. Alternatively, Immel et al. (2025) have developed adaptive precision (AP) energy-mixing schemes coupling classical Embedded Atom Method (EAM) potentials~\cite{daw1993embedded} with ACE~\cite{immel2025adaptive}. While energy-mixing offers a theoretical path to energy conservation, it introduces a spatial switching parameter $\lambda$ whose gradient $\nabla \lambda$ must be computed to account for the forces arising from the changing model assignment~\cite{immel_conservative_2025}.

Attempts to fully learn the routing between model experts in an end-to-end ML framework have also been pioneered by Wood et al.~\cite{wood_uma_2025} through the ``Mixture of Linear Experts'' (MoLE) architecture. They introduce a gating mechanism that determines expert contributions based on global, time-invariant system properties---such as elemental composition, total charge, and the target DFT functional---rather than local atomic environments. This design choice allows the sparse expert weights to be linearly pre-merged into a single effective model prior to inference, successfully decoupling the model's total training capacity (billions of parameters) from its run-time evaluation cost. However, because the routing is global, the resulting potential applies the same effective weights to every atom in the simulation domain.
Building on this line of work, Liu et al.~\cite{liu2026scaling} recently extended end-to-end MoE routing from the configuration level to the atomic level within the DPA3 architecture~\cite{zhang2025dpa3}, demonstrating that element-wise gating based on atomic species, combined with sparse nonlinear expert activation and shared experts, outperforms both dense parameter scaling and the linear MoLE alternative on various benchmarks. While these end-to-end MoE schemes successfully tackle two complementary axes of generalization, i.e. disparate chemical spaces in MoLE (e.g., small molecules versus bulk materials) and chemical specialization across elements in DPA3, neither addresses the spatial efficiency bottleneck in large heterogeneous simulations: in both cases the (activated) expert evaluation cost is incurred uniformly across all atoms, so atoms in simple and homogeneous environments (e.g. bulk metals) are computationally treated with the same high fidelity as those in chemically complex environments (e.g. reactive surface sites). These frameworks therefore optimize representational capacity, per-configuration or per-element, rather than computational effort per region.

In this work, we introduce a multifidelity framework that synthesizes the strengths of these approaches while preserving a simple implementation with flexible but static model assignments, as explained later. Built entirely within the E(3)-equivariant \textit{Allegro} architecture~\cite{musaelian2023learning, tan_high-performance_2025}, our approach creates a fully learnable, end-to-end ML ecosystem. Energy conservation in our framework follows from deriving forces as analytical gradients of a single, well-defined hybrid Hamiltonian; the static domain decomposition is the implementation choice that keeps this Hamiltonian time-independent and differentiable, yielding conservative forces without auxiliary thermostats or switching-gradient corrections. Crucially, to address the mechanical stability challenges at the model interface, we develop a co-training strategy with agreement constraints: an auxiliary loss term that penalizes per-atom energy and force discrepancies between the two models on a shared set of bulk environments. This forces the independent experts to learn a unified physical description of the bulk lattice, allowing each model to specialize while remaining mutually consistent at their interface. This combination allows us to achieve significant computational speedups in large heterogeneous systems by intelligently localizing model capacity, offering a scalable path for first-principles quality simulations of complex reaction environments.

\section{Methods}

\subsection{Static Domain Decomposition}

A material system can be represented as a geometric graph $\mathcal{G}=(\mathcal{V}, \mathcal{E})$, where nodes $i \in \mathcal{V}$ represent atoms characterized by their atomic numbers $Z_i$ and positions $\mathbf{r}_i \in \mathbb{R}^3$. Edges $(i,j) \in \mathcal{E}$ connect atoms within a radial cutoff $r_c$. To enable multifidelity simulations, we partition the global set of atoms $\mathcal{V}$ into $N$ disjoint subsets (here $N=2$ for simplicity), $\mathcal{V}_{\mathcal{A}}$ and $\mathcal{V}_{\mathcal{B}}$, each assigned to a specific interatomic potential model (Model $A$ and Model $B$, respectively). This assignment is handled via a ``Split-Evaluate-Merge'' operation on the computational graph. In the \textit{Split} phase, the global neighbor list is separated into two subgraphs: $\mathcal{G}_{\mathcal{A}}$ contains all edges $(i, j)$ where the central atom $i \in \mathcal{V}_{\mathcal{A}}$, and the same for $\mathcal{G}_{\mathcal{B}}$. Crucially, the neighbors $j$ in these edges may belong to either set. Atoms from $\mathcal{V}_{\mathcal{B}}$ that appear in the local environment of an atom $i \in \mathcal{V}_{\mathcal{A}}$ act as ``ghost atoms''; they contribute geometric information to the embedding of the center atom $i$ but do not contribute their own atomic energy to the output of Model $A$.

In the \textit{Evaluate} phase, the two models are evaluated in parallel on their respective subgraphs. Model $A$ predicts the edge energies for $\mathcal{G}_{\mathcal{A}}$, and Model $B$ for $\mathcal{G}_{\mathcal{B}}$. Finally, in the \textit{Merge} phase, the total potential energy of the hybrid system is constructed by summing the atomic energy contributions from both domains. Special care must be taken regarding the reference energies of the ghost atoms. Allegro decomposes the total energy of the system into a sum of per-atom energies such that
\begin{equation}
E_\text{total} = \sum_k \sigma_{Z_k} E_k + \mu_{Z_k},
\end{equation}
where $\sigma_{Z}$ and $\mu_{Z}$ are a scale and shift for the per-atom energy dependent on the atom type. The model further decomposes each per-atom energy into a sum of ordered-pair edge energies:
\begin{equation}
E_k = \sum_{j \in \mathcal{N}_k} E_{kj}.
\end{equation}
Thus, when the system's graph is split for evaluation across multiple models, each model will process some atoms which appear only as the neighbor to the atoms of the node subset for that model. Nevertheless, the Allegro model will provide an energy shift $\mu_{Z_j}$ for these neighbor atoms, which would lead to an over-counting of the per-atom energy shift if all models across an interface add this value. We remove the extraneous shift energy by defining 
\begin{equation}
E_\text{ghost} = \sum_l \mu_{Z_l},
\end{equation}
where $l$ runs over all atoms $l \in \mathcal{N_X}, l \notin \mathcal{V}_X, X \in \{A,B\}$, the atoms in the neighbor set of a model but not the node set of the model. The total energy then becomes:

$$
E_{\text{total}}(\mathbf{R}) = 
\sum_{i \in \mathcal{V}_{\mathcal{A}}} E_i^{(A)} + 
\sum_{j \in \mathcal{V}_{\mathcal{B}}} E_j^{(B)} - E_{\text{ghost}}.
$$
This additive construction ensures that the interface between the high-fidelity and low-fidelity regions is handled consistently, with the interaction energy across the boundary mediated by model evaluations on the shared geometric environment.

A fundamental advantage of this static assignment scheme is the preservation of exact energy conservation without the need to include additional switching gradients. In dynamic assignment approaches, where atoms switch models based on their spatial position (e.g., distance from a surface), the potential energy surface $E(\mathbf{R})$ becomes discontinuous or time-dependent. In our framework, the assignment of an atom to set $\mathcal{V}_{\mathcal{A}}$ or $\mathcal{V}_{\mathcal{B}}$ is determined by its atom index at initialization and remains fixed. As a result, the Hamiltonian is well-defined and differentiable. The forces on any atom $k$, including those at the interface, are derived via automatic differentiation of the hybrid total energy, $\mathbf{F}_k = -\nabla_{\mathbf{r}_k} E_{\text{total}}(\mathbf{R})$. This mathematically guarantees that the force field is conservative and that total momentum is conserved, enabling stable integration over nanosecond timescales. 

\subsection{Co-training with Agreement Constraints}

While the static assignment scheme guarantees energy conservation, it does not inherently guarantee physical consistency at the interface. If Model $A$ and Model $B$ are trained independently on different datasets (e.g., Model $A$ on surfaces and Model $B$ on bulk configurations), they often may converge to different local minima for the same underlying material phase (or due to stochastic nature of training, converge to different minima despite the same training data). For instance, they may predict different equilibrium lattice constants ($a_0$) and bulk moduli ($B_0$). When these models are spatially coupled, this ``interface mismatch'' manifests as an artificial stress discontinuity: atoms at the boundary experience a non-physical compression or tension as one model attempts to expand or contract the lattice against the other. In bulk fluids, where atoms are not constrained to lattice sites, the same disagreement instead manifests as an artificial chemical potential gradient between the two model regions, driving unphysical density fluctuations near the interface as the models redistribute mass to satisfy their differing equations of state. To resolve this, we implement a co-training strategy that explicitly encourages physical agreement between the models. Instead of optimizing the models solely on their domain-specific tasks, we introduce a shared ``agreement dataset'' $\mathcal{D}_{\text{agree}}$ consisting of bulk configurations. $\mathcal{D}_{\text{agree}}$ can be either a separate dataset, or a subset of atomic environments from the original dataset containing only atoms that are representative of the bulk region. We then define a composite loss function $\mathcal{L}_{\text{total}}$ that balances task specialization with mutual consistency:
\begin{equation}
    \mathcal{L}_{\text{total}} = \alpha \mathcal{L}_{1} + (1 - \alpha) \mathcal{L}_{2}.
\end{equation}

The first term, $\mathcal{L}_{1}$, is the standard MSE loss against ground-truth DFT labels (energies and forces) for each model on its assigned training data~\cite{musaelian2023learning}:

\begin{equation}
    \begin{split}
        \mathcal{L}_1 &= \frac{\lambda_E}{B}\sum_{b=1}^B (\hat{E}_b - E_b)^2 \\
        &\quad + \frac{\lambda_F}{3B}\sum_{i=1}^{B}\sum_{\xi=1}^{3} \left\| -\frac{\partial \hat{E}}{\partial \mathbf{r}_{i,\xi}} - \mathbf{F}_{i,\xi} \right\|^2,
    \end{split}
\end{equation}
where $B$, $E_b$, $\hat{E}_b$, $F_{i,\xi}$ denote the batch size, batch of true energies, batch of predicted energies, and the force component on atom $i$ in spatial direction $\xi$, respectively. The relative weight of each term in $\mathcal{L}_1$ can be specified by the hyperparameters $\lambda_E$ and $\lambda_F$.

The second term, $\mathcal{L}_{2}$, penalizes discrepancies between the predictions of Model $A$ and Model $B$ when evaluated on the same atomic environments $i$ in $\mathcal{D}_{\text{agree}}$:

\begin{equation}
    \mathcal{L}_{2} = \sum_{i \in \mathcal{D}_{\text{agree}}} \left( \lambda'_E\left| \hat{E}^{A}_i - \hat{E}^{B}_i \right|^2 +  \lambda'_F\left\| \hat{\mathbf{F}}_i^{A} - \hat{\mathbf{F}}_i^{B} \right\|^2 \right).
\end{equation}

The hyperparameter $\alpha \in (0, 1]$ controls the trade-off between accuracy on the specific domains and physical consistency at the interface. By minimizing this joint objective, the models are forced to learn a common energy gauge and mechanical response for the bulk material, significantly reducing the stress gradients that would otherwise arise at the model boundary.

While not explicitly enforced or tested in this work, an additional consequence of this construction is that aligning the per-atom energy predictions of Models $A$ and $B$ on a shared bulk environment also aligns their species-dependent energy gauges (i.e., the per-element shifts $\mu_Z$). The mathematical structure of the agreement loss therefore provides a viable route to ensuring consistent chemical potentials across distinct MLIP models. Although immaterial under static assignment, it becomes essential for any extension to dynamic schemes, where atoms migrating between models would otherwise incur a non-physical energy change purely from the change of label.

\section{Experimental Setup}

To validate the framework, we employ a system composed of CO molecules on a platinum surface as a representative test case for heterogeneous catalysis. Platinum nanoparticles are widely used catalysts for oxidation reactions, where the interaction between the metal surface and adsorbed carbon monoxide (CO) molecules drives significant surface restructuring~\cite{owen2024surface}. This system is ideal for benchmarking a spatially decomposed model because it exhibits a clear separation of complexity: the surface region involves complex adsorbate-metal interactions, charge transfer, and dynamic roughening requiring high-fidelity treatment, while the bulk platinum core maintains a relatively simple, stable face-centered cubic (FCC) lattice structure that can be adequately described by a lower-fidelity model.

\begin{figure*}[ht]
    \centering
    \includegraphics[width=\textwidth]{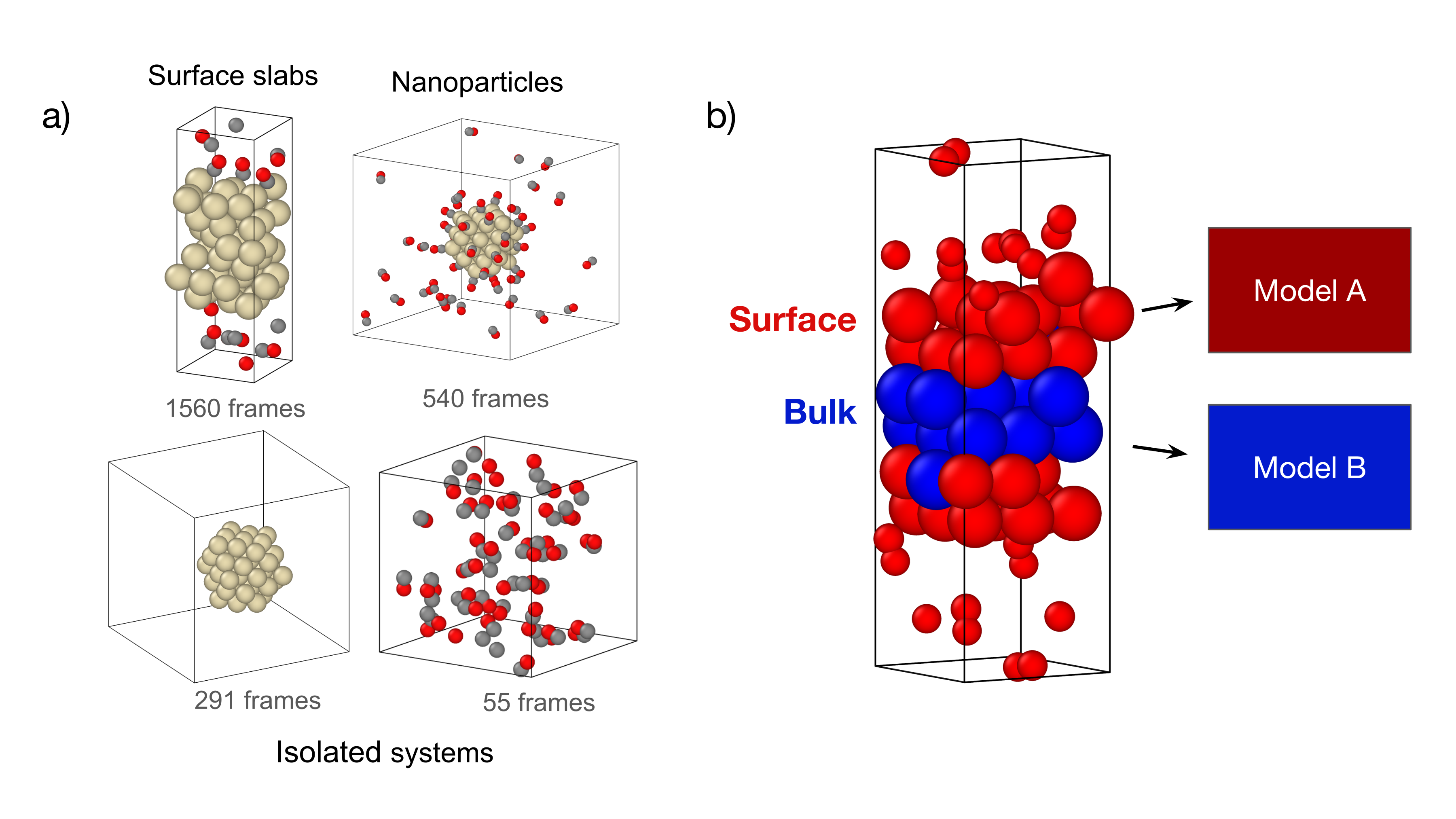}
    \caption{\textbf{Dataset and static domain decomposition.} (a) Representative configurations from the training dataset, which consists of 2,446 DFT frames including surface slabs, free-standing nanoparticles, isolated atomic clusters and CO in gas phase. (b) Illustration of the heuristic assignment strategy used to partition the system. 
    }
    \label{fig:dataset_and_assignment}
\end{figure*}

The training dataset used in this work consists of 2,446 atomic configurations originally generated by Owen et al.~\cite{owen2024surface} using a Bayesian active learning workflow with the FLARE framework~\cite{vandermause2020fly, vandermause2022active}. For the training runs, we employ a 80/10/10\% train/validation/test split. In this work we recalculated the ground-truth energies and forces for the entire dataset using the Vienna Ab-initio Simulation Package (VASP)~\cite{kresse1996efficiency} with the optB88-vdW exchange-correlation functional~\cite{klimes_2010}. The dataset spans a diverse range of environments, including bulk Pt, Pt(111)/Pt(100) slabs with varying CO coverages, and freestanding nanoparticles with a few hundred atoms (see Figure \ref{fig:dataset_and_assignment} a)). This dataset is particularly relevant for computational study because it captures the thermodynamic competition between CO-induced surface roughening and bulk crystallinity, providing a rigorous test for our method's ability to maintain physical consistency at the interface between the surface (high-fidelity) and bulk (low-fidelity) regions.

To partition the atomic systems into the respective domains for Model $A$ and Model $B$, we employed a geometric heuristic based on the local chemical environment, as illustrated in Figure \ref{fig:dataset_and_assignment} b). For each configuration in the dataset, we computed a neighbor list using a cutoff radius of $5.0 \, \text{\AA}$. Any platinum atom possessing a local environment consisting exclusively of other platinum atoms within this radius was classified as ``bulk'' and assigned to the low-fidelity Model $B$. All remaining atoms---including all carbon and oxygen species, as well as platinum atoms with mixed neighborhoods (i.e., those at the surface or interacting with adsorbates)---were assigned to Model $A$. This strategy ensures that the expensive high-fidelity model is concentrated solely where accurate descriptions of charge transfer and surface reconstruction are required.

Additionally, we specify the agreement dataset $\mathcal{D}_{\text{agree}}$ used in the co-training loss function. In this experiment, we did not generate a separate external dataset for this purpose; instead, we utilized the atoms explicitly assigned to the low-fidelity region $\mathcal{B}$ during the static decomposition. For every training frame, the subset of atoms classified as ``bulk'' (i.e., platinum atoms with only platinum neighbors) served simultaneously as the training domain for Model $B$ and the agreement domain for the consistency loss. This ensures that the high-fidelity Model $A$ is forced to match the predictions of Model $B$ specifically in the bulk platinum region.

For the co-training experiment, we instantiate two distinct Allegro models. The high-fidelity model (Model $A$) was constructed with 3 layers, a maximum rotation order of $l_{max}=3$, and 64 tensor features, resulting in approximately 539K trainable parameters. In contrast, the low-fidelity model (Model $B$) was designed for efficiency, using only 2 layers, $l_{max}=1$, and 32 tensor features, totalling roughly 50.7K parameters. The models' capacities and numbers of parameters directly reflect their evaluation speed: the smaller model $B$ is on average 3.8$\times$  faster than model $A$ at evaluating the geometries in this dataset on an A100 80GB GPU.

\section{Results and Discussion}

\subsection{Independent Training vs. Co-training}

\begin{figure*}[ht]
    \centering
    \includegraphics[width=0.9\textwidth]{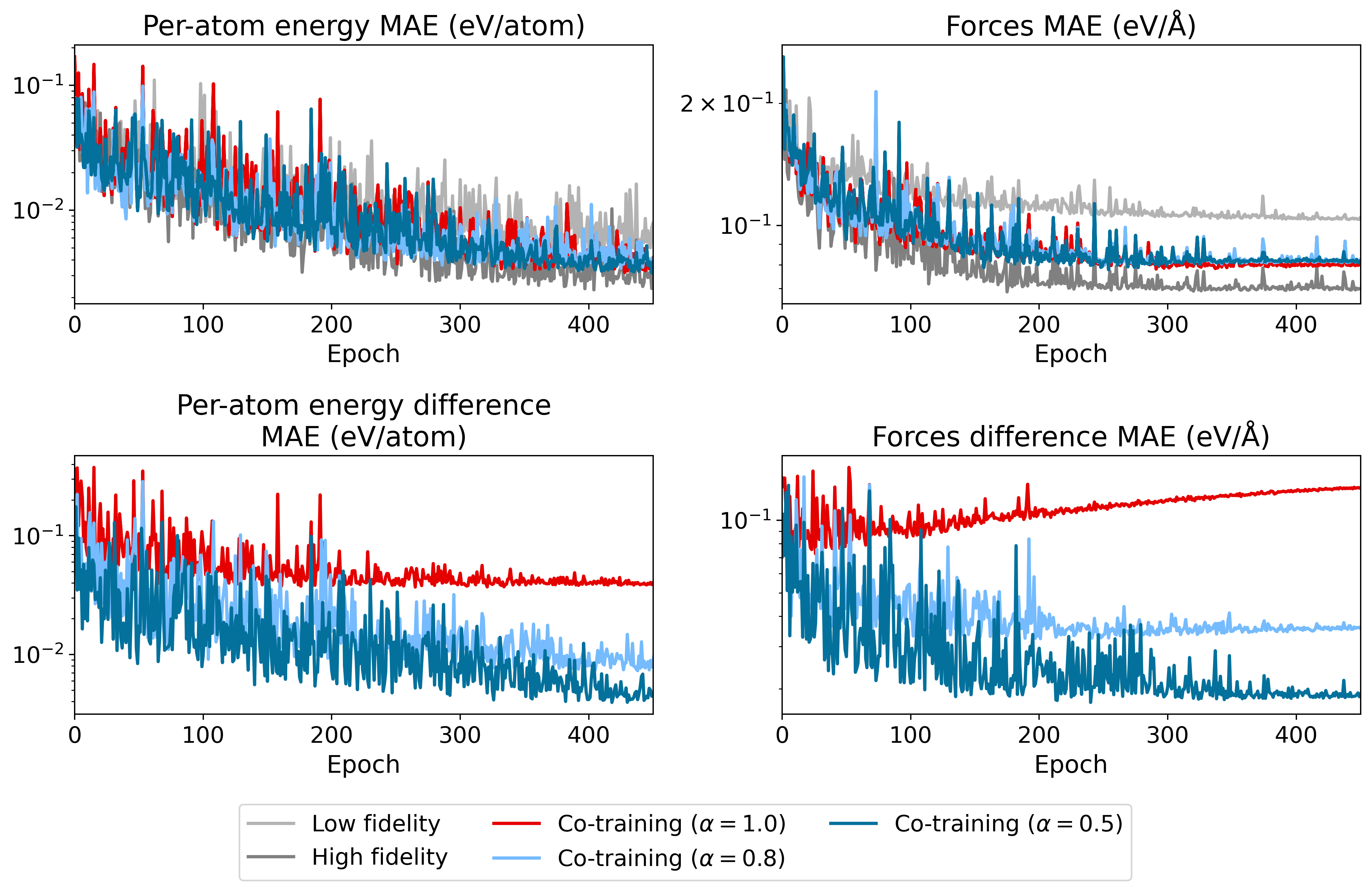}
    \caption{\textbf{Learning dynamics of independent vs. co-trained models.} The top row shows the validation Mean Absolute Error (MAE) against ground truth DFT labels for total energy (left) and atomic forces (right). The co-training schemes (red, blue) achieve an accuracy balance between the pure high-fidelity (gray) and low-fidelity (light gray) baselines. The bottom row displays the discrepancy between the high-fidelity and low-fidelity model predictions on the bulk agreement dataset. Unconstrained co-training (red) leads to significant divergence, while constrained co-training (blue lines) successfully minimizes physical disagreement at the interface. Lower $\alpha$ values (dark blue) impose stronger constraints, resulting in tighter agreement.}
    \label{fig:learning_curves}
\end{figure*}

To evaluate the efficacy of the proposed co-training scheme, we analyze the learning dynamics of the models under different constraint regimes compared to standard independent training. Figure \ref{fig:learning_curves} presents the validation metrics over the course of training for five scenarios: a high-fidelity model trained on the full dataset (gray), a low-fidelity model trained on the full dataset (light gray), co-training with no constraints ($\alpha=1.0$, red), and co-training with two different weighting factors ($\alpha=0.8$, light blue; $\alpha=0.5$, dark blue).

The top row of Figure \ref{fig:learning_curves} illustrates the performance of the combined model on the primary task: predicting the ground-truth DFT energies and forces. The high-fidelity baseline (gray) achieves the lowest error, setting the theoretical upper bound for accuracy given the architecture's capacity. Conversely, the low-fidelity baseline (light gray) exhibits the highest error, reflecting its limited parameter space. Crucially, the co-training schemes (red, blue) achieve a balance of accuracy that lies strictly between these two extremes. The combined multifidelity model performs significantly better than the low-fidelity model alone while approaching the accuracy of the high-fidelity model. This confirms that the framework successfully delegates complex interactions to the high-capacity model while handling bulk regions efficiently, without degrading overall predictive power. Furthermore, the introduction of agreement constraints ($\alpha < 1$) does not negatively impact this task performance compared to unconstrained co-training ($\alpha=1$).

On the other hand, a stark contrast emerges when examining the physical consistency between the models, shown in the bottom row of Figure \ref{fig:learning_curves}. These plots track the mean absolute difference between the predictions of the high-fidelity (Model $A$) and low-fidelity (Model $B$) models evaluated specifically on the bulk agreement dataset $\mathcal{D}_{\text{agree}}$. For the unconstrained case (red line), the disagreement remains high throughout training, and in the case of forces, actually increases (likely due to models' specialization throughout training). This confirms that while independently trained models may both be ``accurate'' relative to their specific training data, they converge to distinct regions of the potential energy landscape for the bulk phase, leading to the mechanical mismatches described in Section 2.

In contrast, the constrained co-training models (blue lines) demonstrate a rapid and sustained reduction in inter-model disagreement. By enforcing the constraint, the optimization process successfully locates a solution where both models predict similar energies and forces for the bulk lattice. We interpret this result through the lens of model capacity: the high-fidelity model, possessing significantly more parameters than the low-fidelity model, effectively utilizes its additional degrees of freedom to align its internal representation of the bulk with that of the simpler model, thereby resolving the interface conflict.

We also observe the impact of the hyperparameter $\alpha$ on the convergence of the agreement metrics. Lower values of $\alpha$ impose a stricter penalty on model's disagreement. Consequently, the run with $\alpha=0.5$ (dark blue) achieves marginally better agreement on energies and forces compared to $\alpha=0.8$ (light blue). While this suggests that stronger constraints yield tighter physical consistency, the optimal value of $\alpha$ is likely system-dependent. A comprehensive hyperparameter sweep to determine the Pareto frontier between accuracy and consistency is left for future work; however, these results sufficiently demonstrate that values in the range of $0.5 - 0.8$ provide a robust solution for the CO/Pt system.

\begin{figure*}[ht]
    \centering
    \includegraphics[width=0.75\textwidth]{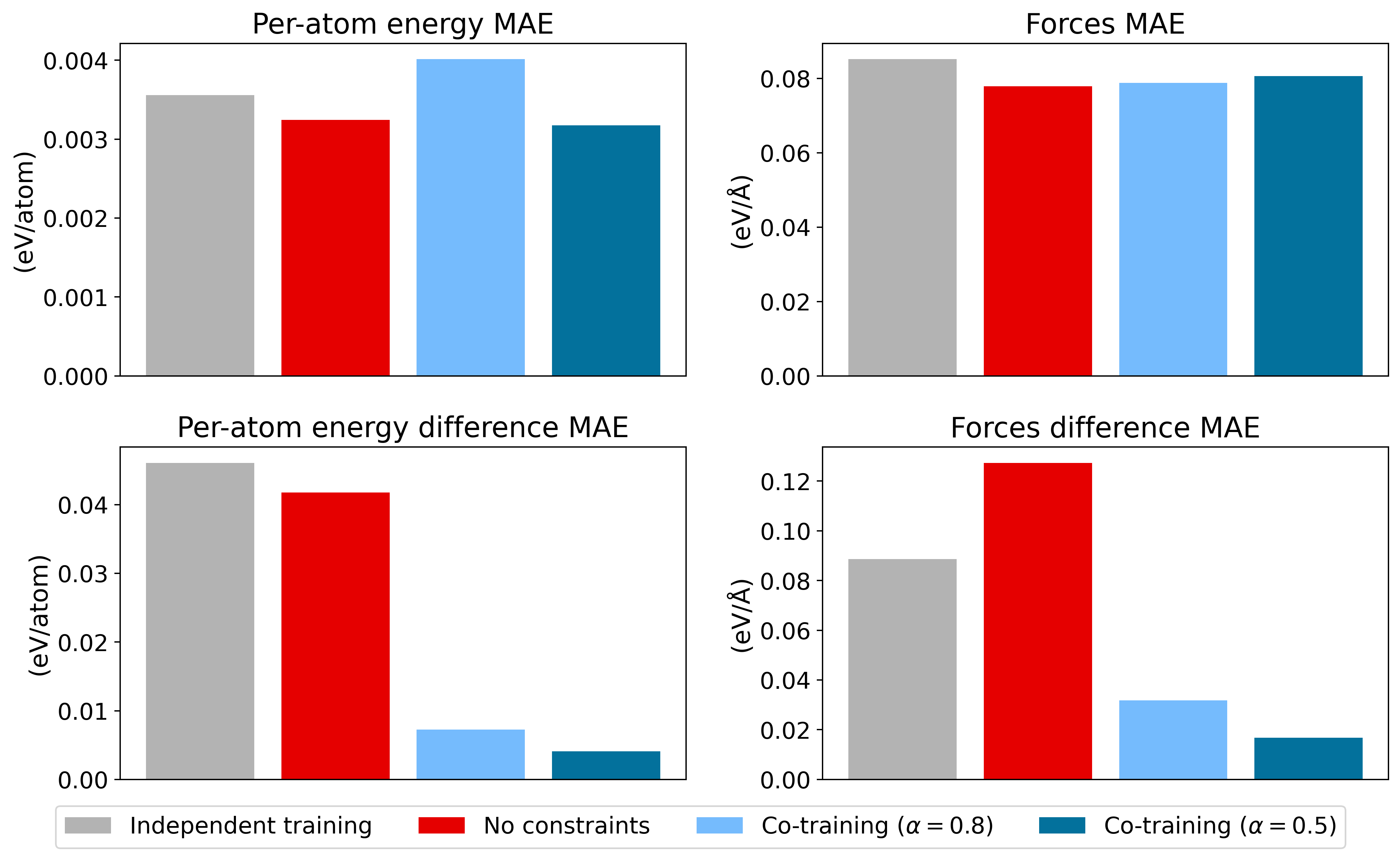}
    \caption{\textbf{Model performance on held-out test set.} Top row: MAE on energies and forces against DFT ground truth. The introduction of agreement constraints ($\alpha < 1$) does not degrade predictive accuracy compared to the unconstrained baseline ($\alpha=1$). Bottom row: Mean absolute difference between high-fidelity and low-fidelity model predictions on bulk test structures. Constrained co-training drastically improves physical consistency at the interface on unseen data.}
    \label{fig:test_set_errors}
\end{figure*}

To rigorously assess generalization, we evaluated the final trained models on a held-out test set comprising 10\% of the original data, which was not used for training or validation. The results, presented in Figure \ref{fig:test_set_errors}, confirm that the trends observed during training hold true for unseen data. The introduction of agreement constraints ($\alpha=0.8, 0.5$) does not increase the MAE on energies or forces significantly relative to the unconstrained case ($\alpha=1.0$); in fact, compared to when both models are trained independently, a slight improvement in energy accuracy is observed, likely due to model specialization. Most importantly, the bottom row of Figure \ref{fig:test_set_errors} demonstrates that the drastic improvement in inter-model agreement is maintained on the test set. The constrained models predict bulk energies and forces that are much more consistent than the unconstrained baseline, ensuring that physical agreement at the interface is a robust, learnable property of the co-training scheme.

\subsection{Model's Interface Stability and Physical Consistency}

\begin{figure*}[ht]
    \centering
    \includegraphics[width=0.75\textwidth]{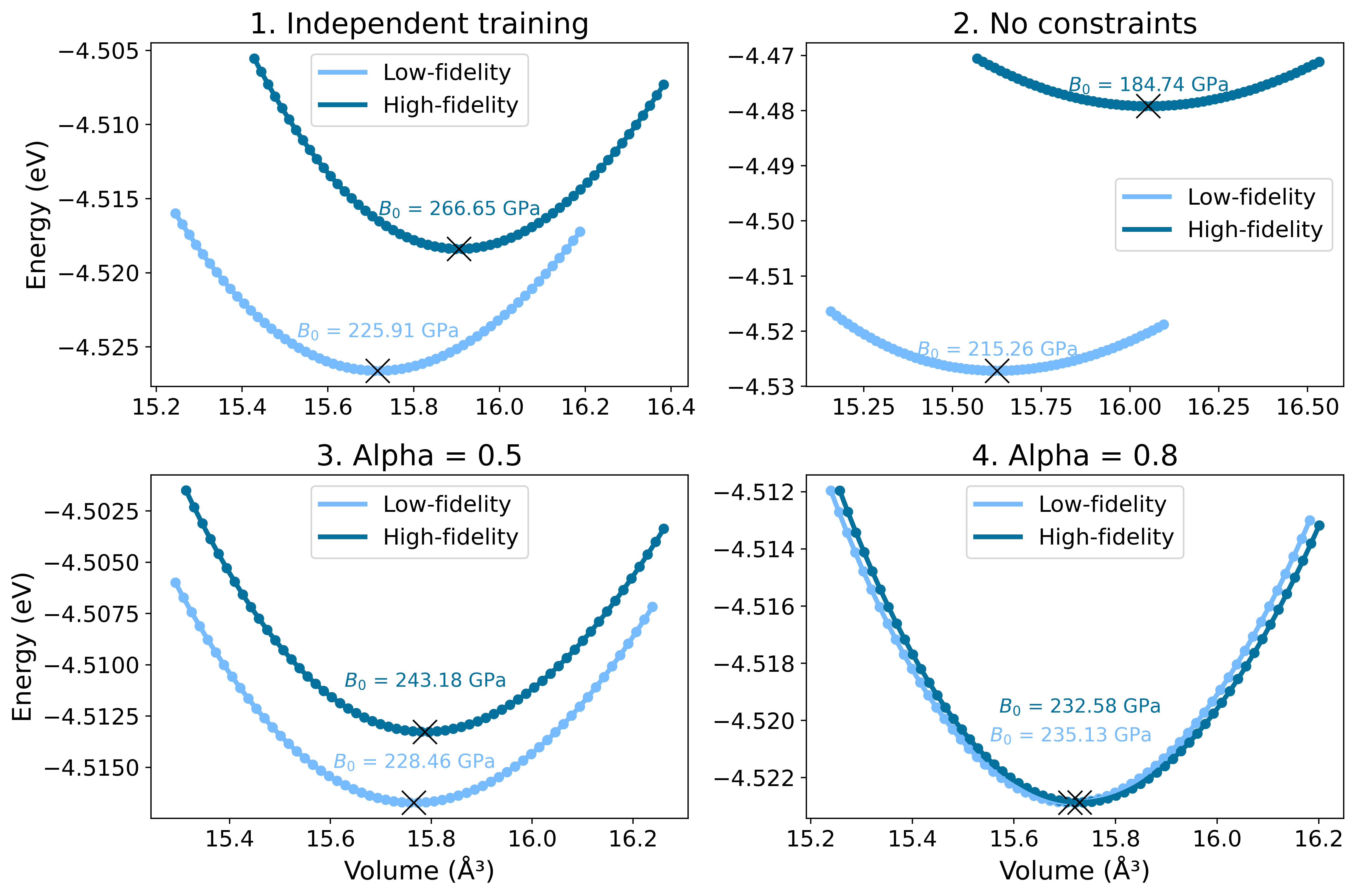}
    \caption{\textbf{Equation of State (EOS) for bulk Platinum predicted by high-fidelity (dark blue) and low-fidelity (light blue) models.} Top row: Models trained independently or co-trained without constraints exhibit significant mismatch in equilibrium volume and bulk modulus ($B_0$), leading to artificial interface stress. Bottom row: Co-training with agreement constraints ($\alpha=0.8, 0.5$) forces the models to learn a consistent physical description of the bulk, aligning the EOS curves and minimizing mechanical mismatch at the interface.}
    \label{fig:eos_comparison}
\end{figure*}

The reduction in energy and force discrepancies observed during training directly translates to improved physical consistency at the model interface. To demonstrate this, we computed the Equation of State (EOS) for bulk platinum using both the high-fidelity and low-fidelity models trained under the different regimes. Figure \ref{fig:eos_comparison} plots the predicted total energy as a function of unit cell volume.

In the case of independent training (top left), the two models converge to significantly different equilibrium states. The low-fidelity model predicts a much smaller bulk modulus ($B_0 \approx 226$ GPa) compared to the high-fidelity model ($B_0 \approx 267$ GPa), along with a noticeable offset in the equilibrium volume. As discussed in Section 2, placing these two models in spatial contact would generate a permanent, artificial stress field at the boundary: the ``softer'' low-fidelity region with lower lattice parameter would be stretched by the ``stiffer'' high-fidelity region, creating a long-ranged stress field on the structure.

When co-trained without constraints ($\alpha=1.0$, top right), this mismatch persists, with the bulk moduli differing by nearly 30 GPa ($185$ vs $215$ GPa). However, the introduction of the agreement constraint ($\alpha < 1$) forces the models to synchronize their mechanical response. For $\alpha=0.8$ (bottom right), the EOS curves align closely, and the predicted bulk moduli come within 3 GPa of each other ($233$ vs $235$ GPa). With a stronger constraint of $\alpha=0.5$ (bottom left), we note that the absolute value of the bulk modulus shifts slightly, despite the energy curves appearing visually well-aligned. This counter-intuitive result suggests a complex interplay between the energy-matching and force-matching terms within the agreement loss. Excessively penalizing point-wise disagreement on specific snapshots may inadvertently constrain the optimization landscape, making it harder for the models to simultaneously match the second-derivative properties (curvature) of the equation of state. This highlights that the relationship between $\alpha$ and physical consistency is likely non-monotonic, reinforcing the need for a more granular hyperparameter sweep and statistical validation across multiple seeds in future work to identify the optimal balance for mechanical compatibility.

\subsection{Computational Efficiency}

\begin{figure*}[ht]
    \centering
    \begin{minipage}{0.2\textwidth}
        \centering
        \includegraphics[width=\textwidth]{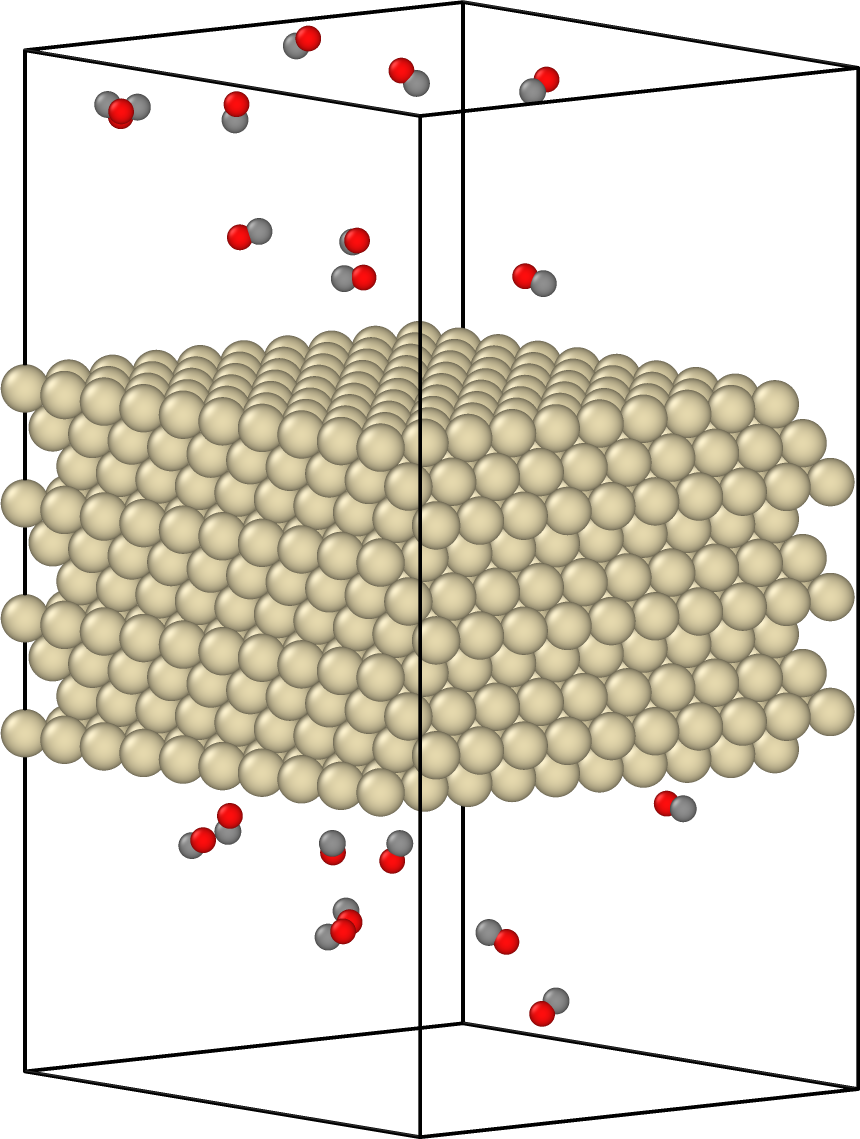}
        \textbf{(a)}
    \end{minipage}
    \begin{minipage}{0.5\textwidth}
        \centering
        \includegraphics[width=\textwidth]{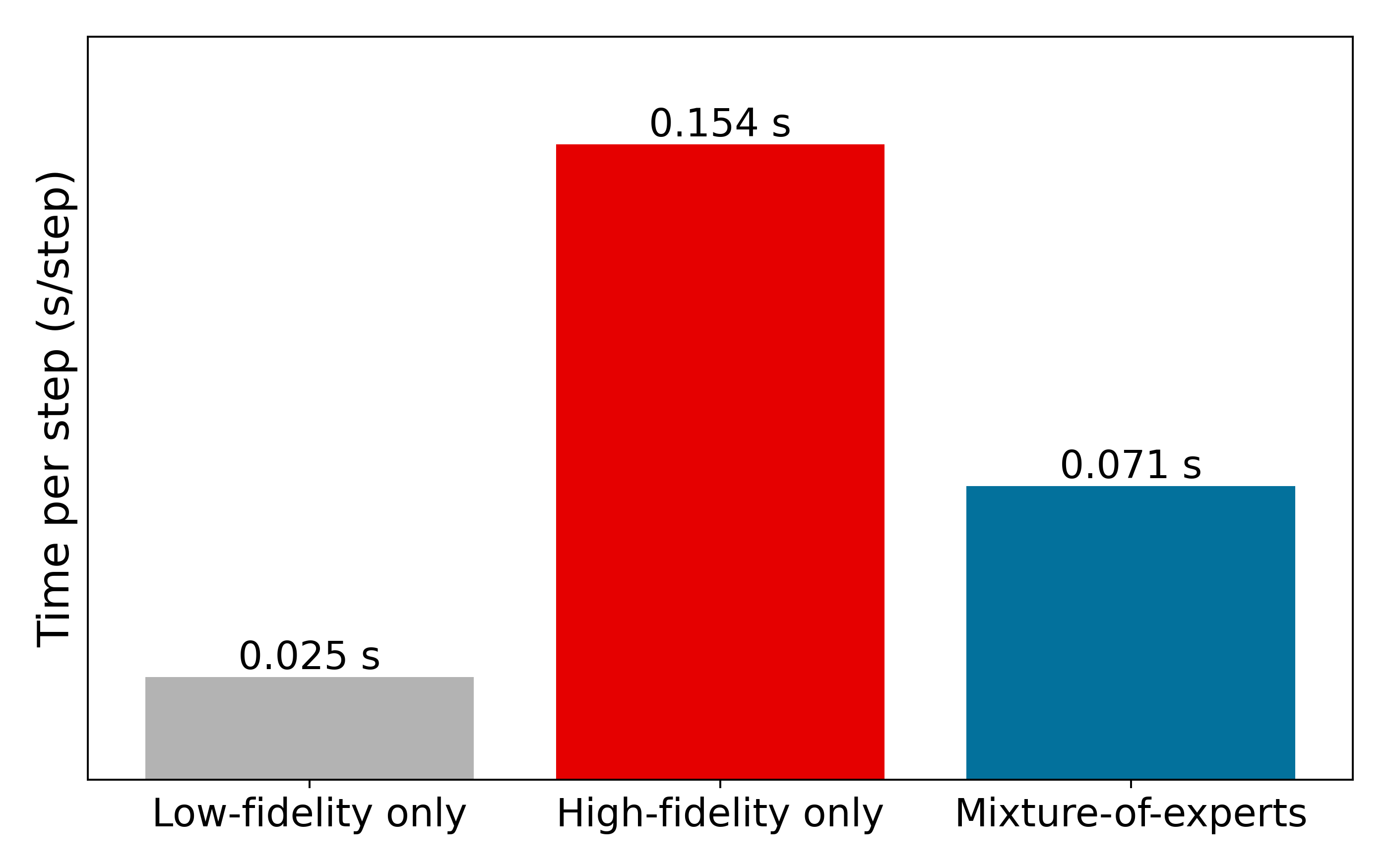}
        \textbf{(b)}
    \end{minipage}
    \caption{\textbf{Computational efficiency benchmark.} (a) The 1,040-atom Pt+CO slab used for MD benchmarking. The high-fidelity region comprises adsorbates and surface Pt layers ($N_{\mathcal{A}}=440$), while the low-fidelity region comprises the bulk Pt core ($N_{\mathcal{B}}=600$). (b) Average wall-clock time per MD timestep for pure low-fidelity, pure high-fidelity, and the combined Multifidelity (Mixture-of-Experts) scheme.}
    \label{fig:efficiency_benchmark}
\end{figure*}

To rigorously evaluate the computational gains offered by our multifidelity approach in a realistic simulation scenario, we constructed a larger-scale test system representative of a catalytic interface. As illustrated in Figure \ref{fig:efficiency_benchmark}(a), the system consists of a thick platinum slab with carbon monoxide molecules adsorbed onto both top and bottom surfaces, totaling 1,040 atoms. This structure is significantly larger than those used in training, allowing us to assess the scaling behavior of the method. Due to the number of atoms, performing molecular dynamics with ab-initio methods such as DFT in this system becomes unfeasible, as the time required per simulation step would surpass many hours.

We applied our static domain decomposition strategy to partition this system based on chemical reactivity. The high-fidelity region ($\mathcal{A}$) was defined to include all adsorbate atoms (C and O) as well as the first two atomic layers of the Pt slab on both surfaces, where fundamental catalytic processes occur. This resulted in 440 atoms assigned to the expensive Model $A$. The remaining 600 atoms, forming the stable bulk core of the slab, were assigned to the low-fidelity region ($\mathcal{B}$) and treated by the efficient Model $B$. We then performed NVE molecular dynamics simulations for 2,000 timesteps using ASE~\cite{ase} to measure the average wall-clock time required per integration step. We compared three different model configurations: (i) a simulation using exclusively the full low-fidelity model, establishing the lower bound for computational cost; (ii) a simulation using exclusively the full high-fidelity model, representing the standard approach that one would traditionally use; and (iii) our proposed Multifidelity (Mixture-of-Experts) scheme using models co-trained with $\alpha=0.5$.

To quantify the efficiency of our implementation, we analyzed the computational cost per edge interaction. The simulation domain contains 51,404 edges in the low-fidelity region ($\mathcal{B}$) and 25,618 edges in the high-fidelity region ($\mathcal{A}$). By normalizing the inference times of the independent baselines by their total edge counts (77,022 edges), we calculated the theoretical minimum time required for a hybrid timestep---assuming zero overhead---to be 0.0678 seconds. Our measured average timestep for the multifidelity simulation is 0.0711 seconds, which corresponds to an efficiency of 95\% relative to the theoretical limit. The remaining 5\% overhead is likely attributable to the operational costs inherent to the static decomposition scheme: the graph manipulation operations required to split the global neighbor list and merge the resulting atomic energies at every step, as well as the latency associated with managing two distinct model graphs in GPU memory. We note that these edge counts represent an instantaneous snapshot of the system's connectivity; during dynamics the partition between high- and low-fidelity edges fluctuates with each neighbor-list rebuild, and is expected to drift modestly toward the high-fidelity side as CO molecules adsorb onto the Pt surface. The reported efficiency should therefore be interpreted as a representative estimate rather than a strict bound.


Comparing the absolute wall-clock times, the traditional approach of treating the entire system with the high-fidelity model required significantly higher computational resources. By restricting the expensive model solely to the chemically active interface, the Multifidelity framework achieved a speedup of more than $2\times$ relative to the pure high-fidelity baseline. This performance gain is a direct consequence of the static domain decomposition, which allowed the efficient low-fidelity model to handle the majority of interactions (approximately $67\%$ of the total edges) in the bulk, effectively concentrating the computational budget where it provides the most value.

\section{Conclusion}

In this work, we have presented a robust multifidelity framework for atomistic simulations that leverages the strict locality of E(3)-equivariant graph neural networks to bridge the gap between predictive accuracy and computational scalability. By partitioning the simulation domain into high-fidelity and low-fidelity regions based on chemical complexity, our approach allows for the efficient allocation of computational resources without sacrificing physical rigor.

Our static domain decomposition scheme, implemented within the Allegro architecture, guarantees exact energy conservation and conservative forces, a critical advantage over previous force-mixing methods that produce non-conservative forces. Furthermore, we addressed the challenge of mechanical mismatch at the model interface through a co-training strategy with agreement constraints. We demonstrated that by penalizing discrepancies between the models on a shared bulk dataset, we can force independently parameterized potentials to learn a consistent physical description of the material, effectively eliminating artificial interface stresses.

Benchmarks on a realistic catalytic system (Pt+CO) confirm that the proposed Mixture-of-Experts model achieves predictive accuracy comparable to a full high-fidelity simulation while delivering a computational speedup of over $2\times$. This establishes our framework as a viable path for scaling first-principles-quality simulations to experimentally relevant length and time scales. Finally, our formulation points to a viable route to ensuring consistent chemical potentials across distinct MLIP models, which is critical for chemically meaningful simulations treated with dynamic assignment schemes. Future work will focus on extending this methodology to include dynamic assignment schemes that can adaptively refine the high-fidelity region during the course of a simulation in a fully differentiable way.

\section{Code availability}

The machine learning framework used is built around the open-source ecosystem of NequIP \url{github.com/mir-group/nequip}. The specific architecture of Allegro is implemented in a separate code available at \url{github.com/mir-group/allegro}.


The methods developed in this work for co-training models and handling atomic graphs were implemented in an additional repository (currently private) that will soon be available as an open-source NequIP extension, upon publication of this work's results. Please contact the author at \url{demirand@mit.edu} for early access to the code.

\begin{acknowledgments}
Work at Harvard University by G.M.N. and B.K. was supported by Bosch Research. M.L.D. acknowledges support from the Harvard University Materials Research Science and Engineering Center funded by the National Science Foundation grant DMR-2011754. L.Z. acknowledges support from the National Science Foundation Graduate Research Fellowship under Grant No.~DGE 2140743. The computations in this paper were run on the FASRC Cannon cluster supported by the FAS Division of Science Research Computing Group at Harvard University.
\end{acknowledgments}

\bibliographystyle{unsrt} 
\bibliography{main}       

\begin{thebibliography}{10}

\bibitem{BatatiaEtAlDesignSpaceE3Equivariant2022}
Ilyes Batatia, Simon Batzner, Dávid~Péter Kovács, Albert Musaelian, Gregor
  N.~C. Simm, Ralf Drautz, Christoph Ortner, Boris Kozinsky, and Gábor
  Csányi.
\newblock The {Design} {Space} of {E}(3)-{Equivariant} {Atom}-{Centered}
  {Interatomic} {Potentials}, November 2022.
\newblock arXiv:2205.06643 [stat].

\bibitem{batzner20223}
Simon Batzner, Albert Musaelian, Lixin Sun, Mario Geiger, Jonathan~P Mailoa,
  Mordechai Kornbluth, Nicola Molinari, Tess~E Smidt, and Boris Kozinsky.
\newblock E (3)-equivariant graph neural networks for data-efficient and
  accurate interatomic potentials.
\newblock {\em Nature communications}, 13(1):2453, 2022.

\bibitem{NEURIPS2022_4a36c3c5}
Ilyes Batatia, David~P Kovacs, Gregor Simm, Christoph Ortner, and Gabor Csanyi.
\newblock Mace: Higher order equivariant message passing neural networks for
  fast and accurate force fields.
\newblock In S.~Koyejo, S.~Mohamed, A.~Agarwal, D.~Belgrave, K.~Cho, and A.~Oh,
  editors, {\em Advances in Neural Information Processing Systems}, volume~35,
  pages 11423--11436. Curran Associates, Inc., 2022.

\bibitem{bochkarev2024graph}
Anton Bochkarev, Yury Lysogorskiy, and Ralf Drautz.
\newblock Graph atomic cluster expansion for semilocal interactions beyond
  equivariant message passing.
\newblock {\em Physical Review X}, 14(2):021036, 2024.

\bibitem{QMMM}
A.~Warshel and M.~Levitt.
\newblock Theoretical studies of enzymic reactions: Dielectric, electrostatic
  and steric stabilization of the carbonium ion in the reaction of lysozyme.
\newblock {\em Journal of Molecular Biology}, 103(2):227--249, 1976.

\bibitem{bernstein2009hybrid}
Noam Bernstein, James~R Kermode, and Gabor Csanyi.
\newblock Hybrid atomistic simulation methods for materials systems.
\newblock {\em Reports on Progress in Physics}, 72(2):026501, 2009.

\bibitem{sun2024mixture}
Manxi Sun, Wei Liu, Jian Luan, Pengzhi Gao, and Bin Wang.
\newblock Mixture of diverse size experts.
\newblock In Franck Dernoncourt, Daniel Preo{\c{t}}iuc-Pietro, and Anastasia
  Shimorina, editors, {\em Proceedings of the 2024 Conference on Empirical
  Methods in Natural Language Processing: Industry Track}, pages 1608--1621,
  Miami, Florida, US, November 2024. Association for Computational Linguistics.

\bibitem{wang2025hmoe}
An~Wang, Xingwu Sun, Ruobing Xie, Shuaipeng Li, Jiaqi Zhu, Zhen Yang, Pinxue
  Zhao, Weidong Han, Zhanhui Kang, Di~Wang, et~al.
\newblock Hmoe: Heterogeneous mixture of experts for language modeling.
\newblock In {\em Proceedings of the 2025 Conference on Empirical Methods in
  Natural Language Processing}, pages 21954--21968, 2025.

\bibitem{birks2025efficient}
Fraser Birks, Matthew Nutter, Thomas~D. Swinburne, and James~R. Kermode.
\newblock Efficient and accurate spatial mixing of machine learned interatomic
  potentials for materials science.
\newblock {\em npj Computational Materials}, 12(1), February 2026.

\bibitem{LysogorskiyEtAlPerformantImplementationAtomic2021}
Yury Lysogorskiy, Cas van~der Oord, Anton Bochkarev, Sarath Menon, Matteo
  Rinaldi, Thomas Hammerschmidt, Matous Mrovec, Aidan Thompson, Gábor Csányi,
  Christoph Ortner, and Ralf Drautz.
\newblock Performant implementation of the atomic cluster expansion ({PACE})
  and application to copper and silicon.
\newblock {\em npj Computational Materials}, 7(1):97, June 2021.

\bibitem{XieEtAlUltrafastInterpretableMachinelearning2023}
Stephen~R. Xie, Matthias Rupp, and Richard~G. Hennig.
\newblock Ultra-fast interpretable machine-learning potentials.
\newblock {\em npj Computational Materials}, 9(1):162, September 2023.

\bibitem{daw1993embedded}
Murray~S Daw, Stephen~M Foiles, and Michael~I Baskes.
\newblock The embedded-atom method: a review of theory and applications.
\newblock {\em Materials Science Reports}, 9(7-8):251--310, 1993.

\bibitem{immel2025adaptive}
David Immel, Ralf Drautz, and Godehard Sutmann.
\newblock Adaptive-precision potentials for large-scale atomistic simulations.
\newblock {\em The Journal of Chemical Physics}, 162(11), 2025.

\bibitem{immel_conservative_2025}
David Immel, Ralf Drautz, and Godehard Sutmann.
\newblock Conservative adaptive-precision interatomic potentials, December
  2025.
\newblock arXiv:2512.07693 [physics].

\bibitem{wood_uma_2025}
Brandon~M Wood, Misko Dzamba, Xiang Fu, Meng Gao, Muhammed Shuaibi, Luis
  Barroso-Luque, Kareem Abdelmaqsoud, Vahe Gharakhanyan, John~R. Kitchin,
  Daniel~S. Levine, Kyle Michel, Anuroop Sriram, Taco Cohen, Abhishek Das,
  Sushree~Jagriti Sahoo, Ammar Rizvi, Zachary~Ward Ulissi, and C.~Lawrence
  Zitnick.
\newblock {UMA}: A family of universal models for atoms.
\newblock In {\em The Thirty-ninth Annual Conference on Neural Information
  Processing Systems}, 2026.

\bibitem{liu2026scaling}
Yuzhi Liu, Duo Zhang, Anyang Peng, Weinan E, Linfeng Zhang, and Han Wang.
\newblock Scaling machine learning interatomic potentials with mixtures of
  experts.
\newblock {\em arXiv preprint arXiv:2603.07977}, 2026.

\bibitem{zhang2025dpa3}
Duo Zhang, Anyang Peng, Chengqian Cai, Wenshuo Li, Yu~Zhou, Jinzhe Zeng, Mingyu
  Guo, Chengjian Zhang, Bowen Li, Hangrui Jiang, et~al.
\newblock Graph neural network model for the era of large atomistic models.
\newblock {\em arXiv preprint arXiv:2506.01686}, 2025.

\bibitem{musaelian2023learning}
Albert Musaelian, Simon Batzner, Anders Johansson, Lixin Sun, Cameron~J Owen,
  Mordechai Kornbluth, and Boris Kozinsky.
\newblock Learning local equivariant representations for large-scale atomistic
  dynamics.
\newblock {\em Nature Communications}, 14(1):579, 2023.

\bibitem{tan_high-performance_2025}
Chuin~Wei Tan, Marc~L. Descoteaux, Mit Kotak, Gabriel de~Miranda Nascimento,
  Se{\'a}n~R. Kavanagh, Laura Zichi, Menghang Wang, Aadit Saluja, Yizhong~R.
  Hu, Tess Smidt, Anders Johansson, William~C. Witt, Boris Kozinsky, and Albert
  Musaelian.
\newblock High-performance training and inference for deep equivariant
  interatomic potentials.
\newblock {\em Digital Discovery}, 2026.
\newblock Advance Article.

\bibitem{owen2024surface}
Cameron~J Owen, Nicholas Marcella, Christopher~R O'Connor, Taek-Seung Kim,
  Ryuichi Shimogawa, Clare~Yijia Xie, Ralph~G Nuzzo, Anatoly~I Frenkel,
  Christian Reece, and Boris Kozinsky.
\newblock Surface roughening in nanoparticle catalysts.
\newblock {\em arXiv preprint arXiv:2407.13643}, 2024.

\bibitem{vandermause2020fly}
Jonathan Vandermause, Steven~B Torrisi, Simon Batzner, Yu~Xie, Lixin Sun,
  Alexie~M Kolpak, and Boris Kozinsky.
\newblock On-the-fly active learning of interpretable bayesian force fields for
  atomistic rare events.
\newblock {\em npj Computational Materials}, 6(1):20, 2020.

\bibitem{vandermause2022active}
Jonathan Vandermause, Yu~Xie, Jin~Soo Lim, Cameron~J Owen, and Boris Kozinsky.
\newblock Active learning of reactive bayesian force fields applied to
  heterogeneous catalysis dynamics of h/pt.
\newblock {\em Nature Communications}, 13(1):5183, 2022.

\bibitem{kresse1996efficiency}
Georg Kresse and J{\"u}rgen Furthm{\"u}ller.
\newblock Efficiency of ab-initio total energy calculations for metals and
  semiconductors using a plane-wave basis set.
\newblock {\em Computational materials science}, 6(1):15--50, 1996.

\bibitem{klimes_2010}
Jiří Klimeš, David~R Bowler, and Angelos Michaelides.
\newblock Chemical accuracy for the van der waals density functional.
\newblock {\em Journal of Physics: Condensed Matter}, 22(2):022201, dec 2009.

\bibitem{ase}
Ask Hjorth~Larsen, Jens J{\o}rgen~Mortensen, Jakob Blomqvist, Ivano~E Castelli,
  Rune Christensen, Marcin Du{\l}ak, Jesper Friis, Michael~N Groves, Bj{\o}rk
  Hammer, Cory Hargus, et~al.
\newblock The atomic simulation environment—a python library for working with
  atoms.
\newblock {\em Journal of Physics: Condensed Matter}, 29(27):273002, 2017.

\end{thebibliography}

\end{document}